\documentclass[prl,twocolumn,aps,amsmath,amssymb,superscriptaddress,showpacs]{revtex4-1}

\usepackage{rotating}
\usepackage{amsmath}
\usepackage{amssymb}
\usepackage{comment}
\usepackage{color}
\usepackage{graphicx}
\usepackage[active]{srcltx}
\usepackage[sort&compress]{natbib}
\usepackage{amssymb}
\usepackage[dvips]{epsfig}
\usepackage{float}
\usepackage{epsfig}

\renewcommand{\[}{\left[}
\renewcommand{\]}{\right]}
\renewcommand{\(}{\left(}
\renewcommand{\)}{\right)}

\newcommand{\be}{\begin{equation}}
\newcommand{\ee}{\end{equation}}
\newcommand{\bea}{\begin{eqnarray}}
\newcommand{\eea}{\end{eqnarray}}

\def\ai		{{\em ab--initio}}

\def\rr		{{\bf r}}

\def\HH		{{\bf H}}
\def\UU		{{\bf U}}
\def\EE		{{\bf E}}

\def\kk		{{\bf k}}

\def\gc         {\gamma}

\def\gee        {\epsilon}

\renewcommand{\[}{\left[}
\renewcommand{\]}{\right]}
\renewcommand{\(}{\left(}
\renewcommand{\)}{\right)}

\def\dk         {\frac{d\,{\bf k}}{\(2 \pi\)^3}}
\def\dk1        {\frac{d\,{\bf k}_1}{\(2 \pi\)^3}}

\def\dk         {\frac{d\,{\bf k}}{\(2 \pi\)^3}}

\def\bverb      {\renewcommand{\baselinestretch}{1}\begin{verbatim}}
\def\everb  	{\end{verbatim}\renewcommand{\baselinestretch}{1.3}}

\newcommand{\cnr} {Istituto di Struttura della Materia of the National Research Council, Via Salaria Km 29.3, I-00016 Montelibretti, Italy}
\newcommand{\etsf} {European Theoretical Spectroscopy Facilities (ETSF)}

\begin{document}
 
\title{Ultra--fast carriers relaxation in bulk silicon following photo--excitation with a short and polarized laser pulse}

\author{Davide Sangalli}
\affiliation{\cnr} 
\affiliation{\etsf} 

\author{Andrea Marini}
\affiliation{\cnr} 
\affiliation{\etsf} 

\date{\today}
\begin{abstract}
A novel approach based on the merging of the out--of--equilibrium Green's function method with the \ai\,
Density--Functional--Theory is used to describe
the ultra--fast carriers relaxation in Silicon. The results are compared with recent two photon photo--emission
measurements. We show that the interpretation of the carrier relaxation
in terms of $L\rightarrow X$ inter--valley scattering is not correct.
The ultra--fast dynamics measured experimentally is, instead,
due to the scattering between degenerate $L$ states that is activated by the non symmetric population of the
conduction bands induced by the laser field. This ultra--fast relaxation is, then, entirely due
to the specific experimental setup and  it can be interpreted by introducing a novel definition of the quasi--particle lifetimes
in an out--of--equilibrium context.
\end{abstract} 





\maketitle

Silicon ($Si$) is a fundamental building block of semiconductors physics and microelectronics industry~\cite{Omara1}.
The miniaturization of $Si$--based devices to the nano--scale regime and the never ending search for
faster devices call for a deep understanding of the fundamental quantum--mechanical process
that governs the ultra--short time dynamics of electrons and holes~\cite{Haug1,Sundaram1}.
Most of the knowledge of the electronic and optical properties of $Si$ remain, however, limited to
the equilibrium regime. Only recently
the development of ultra--fast laser pulses~\cite{Cerullo1,Brida1} has opened the opportunity to directly investigate
the real--time dynamics in the non--equilibrium (NEQ) regime~\cite{Shah1}.

In real--time experiments the system is initially perturbed with a short laser pulse (the pump) followed by a 
second weaker pulse (the probe) that measures a specific physical observable like,
for example, the absorption~\cite{Hase1,Hase2}
or the photo--emission~\cite{Ichibayashi1,Ichibayashi2} spectra. 
The dynamics induced by the pump is, then, monitored by observing and analyzing
the modifications induced in these observable by the presence of photo--excited carriers.

Despite the enormous experimental interest and the continuous development of more refined
experimental techniques, the simulation methods are still 
based either on equilibrium first--principles approaches 
or on NEQ model Hamiltonians. 

In the case of model Hamiltonians the relaxation paths
can be calculated by using the non--equilibrium Green's function\,(NEGF)~\cite{Haug1,Kwong1}
or the Monte Carlo~\cite{Jacoboni1} methods. The advantage of these approaches is that the modifications
induced by the presence of photo--excited charges is correctly taken into account in
the evaluation of the scattering transitions. However
 {\em ad--hoc} parameters must be introduced to describe both the
photo--excitation process and the specific material properties.

First--principles simulations are commonly performed by using
time--dependent Density--Functional Theory~\cite{K.1996}
or equilibrium Many--Body Perturbation Theory~\cite{Onida1}. In the first case 
the coupling with the laser pulse is described but the dissipative processes
are neglected~\cite{Sato1,Sato2} or described in an empirical way~\cite{Husser1}.
In the second case the laser pulse is
replaced by some ad--hoc initial guess of the carriers distribution and, as a consequence, the
scattering rates are derived from the equilibrium and static quasi--particle\,(QP) lifetimes~\cite{Bernardi1}.

In this Letter we demonstrate that only by a careful and \ai\, description of both the
photo--excitation process and the {\em full time dependence} of the non--equilibrium
carrier scatterings it is possible to device a successful, parameter--free, accurate and predictive approach to the
interpretation of real--time experiments.
This goal will be achieved by merging the atomistic description of the \ai\, approach with the accuracy of the NEGF method.

In particular we will 
reproduce the time evolution carriers in bulk $Si$, observed in a recent
2PPE~\cite{Ichibayashi1,Ichibayashi2} experiment, without relying on any parameter. 
We will highlight and discuss the different scattering channels created by the 
pump excitation showing
the existence of two different decay regimes: an ultra--fast regime due to transitions between
energetically degenerated states, made possible by the symmetry breaking caused by the pump pulse; and a slower
regime, where the carriers are taken to the minimum of the conduction band.
In addition we will investigate the very fundamental problem of defining the lifetime of a photo--excited carrier.
We will show that this definition differs from the equilibrium case, also at a very low carriers concentration.

Our theoretical framework is based on the NEGF theory which describes the time evolution of the 
lesser Green's function $G_{nm\kk}^<\(t\)$~\cite{Haug1}.
$G_{nm\kk}^<\(t\)$ is a matrix in the band indexes ($n$ and $m$) while it is diagonal in the $\kk$--points.
It is solution of the Baym--Kadanoff equations\,(BKE)~\cite{Haug1} which,
projected in the Kohn--Sham (KS) basis set~\cite{Attaccalite1,Marini1},
describes at the same time the time--dependent polarization (via the off--diagonal matrix elements) and the carrier dynamics. 

Indeed restricting to the diagonal matrix elements of $G^<$ we have access to the time--dependent
occupations of the electronic levels defined as $f_{n\kk}\(t\)\equiv \Im\[G^<_{nn\kk}\(t\)\]$. 
In this case the BKE
reduces~\cite{Marini1} to a simple non--linear equation:
\begin{align} 
\partial_t f_{n\kk}\(t\) = \left.\partial_t f_{n\kk}\(t\)\right|_{pump}+\left.\partial_t f_{n\kk}\(t\)\right|_{relax},
\label{eq:1}
\end{align}
with 
\begin{gather} 
\left.\partial_t f_{n\kk}\(t\)\right|_{pump}=\[\HH_{EQ}+{\bf\Delta\Sigma}_{Hxc}(t)+\UU(t),G^<(t) \]_{nn\kk},
\label{eq:2}\\
\left.\partial_t f_{n\kk}\(t\)\right|_{relax}=-\gc^{(e)}_{n\kk}(t) f^{(e)}_{n\kk}(t)+\gc^{(h)}_{n\kk}(t) f^{(h)}_{n\kk}(t).
\label{eq:3}
\end{gather}
Here ${f^{(e)}_{n\kk}=f_{n\kk}}$ and ${f^{(h)}_{n\kk}=1-f_{n\kk}}$
are the electron and the hole occupations.
All ingredients of Eqs.\ref{eq:2}--\ref{eq:3} are calculated \ai\,\cite{yambo_code,Giannozzi2009} and the formal
definition of the $\gc^{\(e,h\)}$ lifetime is done by using the NEGF theory~\cite{Marini1}. 
${\HH_{EQ}=H_{KS}+\Delta E^{QP}}$ is the KS
Hamiltonian~\cite{R.M.Dreizler1990} ($H_{KS}$) corrected by equilibrium many--body effects ($\Delta E^{QP}$). 
$H_{KS}$ describes the atomic structure of the system, providing the \ai\, basis of the simulation
and making this approach universal and applicable to any kind of material~\cite{Onida1}. 
${{\bf\Delta\Sigma}_{Hxc}\(t\)}$ is
the NEQ Hartree plus $COHSEX$ self--energy and ensures a correct coupling with the laser pulse.
Indeed the ${{\bf\Delta\Sigma}_{Hxc}\(t\)}$ structure is such to reduce the BKE,
in the linear regime, to the well--known Bethe--Salpeter equation~\cite{Attaccalite1}.

$\UU$ and $\gc^{\(e,h\)}$ represent the most important building blocks of our approach. The $\UU$ 
operator describes, in the length gauge, the coupling with  
external pump field $\mathbf E\(t\)$: 
$\hat{U}\(t\) = - \hat{\rr} \cdot \EE_{pump}\(t\)$.

${\gc^{\(e\)}}$ and ${\gc^{\(h\)}}$ are the electron and the hole lifetimes respectively and describe
relaxation and dissipation of the photo--excited carriers~\cite{Marini1,Sangalli1}.
$\gc^{\(e,h\)}$ comprises contributions from electron--electron\,(e--e) and electron--phonon\,(e--p) scatterings.
In the present case
both  ${\gc^{\(e\)}}$ and ${\gc^{\(h\)}}$ are time--dependent and 
non--linear functions of the occupations $f_{n\kk}\(t\)$. 

Eq.\ref{eq:3} makes clear the different role played by ${\gc^{\(e\)}}$ and ${\gc^{\(h\)}}$. 
They describe the elemental process
where an initial electron(hole) is scattered in another electron(hole) emitting or absorbing an 
electron--hole pair (e--e channel)~\cite{Sundaram1} or a phonon
(e--p channel)~\cite{Sundaram1,DasSarma1,Allen1}.
In the e--p case the energy is transferred back and forth from the electronic to the phonon sub--systems
until a thermal equilibrium is reached~\cite{Haug1}.
Thus Eq.\ref{eq:1} describes both relaxation and dissipation.
In the $(e)$ channel the term ${\gc^{\(e\)}f^{(e)}}$ describes the removal of electrons
from the state $(n\kk)$ and gives a negative contribution to $\partial_t f^{(e)}_{n\kk}\(t\)$,
while ${\gc^{\(h\)}f^{(h)}}$ describes removal of holes, thus the filling of the state $(n\kk)$,
and gives a positive contribution.

\begin{figure}[t]
\centering
\epsfig{figure=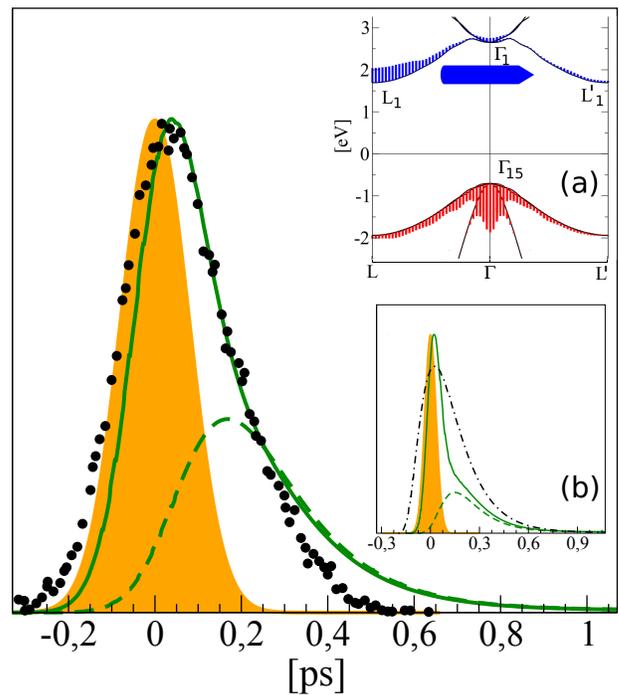, clip=,width=0.45\textwidth}
\caption{\footnotesize{(color online) The time dependent occupation of the $L_1$ (green continuous line) and $L'_1$
(green dashed line) levels are compared with experimental data (black dots) from Ref.~\onlinecite{Ichibayashi1}.
The envelope of the laser pulse is also represented (orange shadow).
In the inset $(a)$ the carriers population (electrons in blue, holes in red) is super imposed on the
band--structure~\cite{NoteBS} at $t=0$.
The blue arrow indicates the direction of the ultra--fast $L_1\rightarrow L'_1$ scattering process.
In the inset $(b)$ the dynamics with a shorter laser pulse ($\sigma=50$\,fs) 
is compared with a {\em gedanken experiment} (black dot--dashed line) where
the same density of carriers is placed {\em by hand} at $t=0$ in the $\Gamma_{15}$ state. With the shorter pulse,
the difference between the fast $L_1\rightarrow L'_1$ scattering and the slower $L_1\rightarrow X_1$ transitions
becomes evident.
}}
\label{fig:1}
\end{figure}

In the 2PPE experiment we aim at describing~\cite{Ichibayashi1,Ichibayashi2}, a $Si$ wafer, oriented both
along the $[111]$ and the $[100]$ surface directions, is excited with a laser pulse at room temperature.
The photo--excited sample is, then, probed with a second laser pulse
that photo--emits in the continuum the excited carriers. The photo--emitted current of electrons
is measured as a function of the time delay between the pump and the probe. The final
result is a measure of the time--dependent occupation of the valence bands (represented by the dots in the main frame of Fig.\ref{fig:1}).
More specifically we consider the population of carriers near the point $L_1$, i.e. at $E\approx 1.6\ eV$
above the Fermi level~\cite{bulk_prop}.

We thus consider the electronic real--time dynamics in bulk $Si$, under the action of a laser pulse,
whose parameters are taken directly from the 2PPE experiment~\cite{Ichibayashi1,Ichibayashi2}. The pulse is
centered at around ${3.4\ eV}$, with duration (the full width at half maximum) of ${\sigma=110\ fs}$ and
intensity of ${10^4\ kW/cm^2}$ corresponding to an electric field intensity of ${6\times 10^6\ V/m}$.
The total fluence is ${4\times 10^{-8}\ KJ/cm^3}$, which means that the pump field creates a carriers density
of about ${4.5\times 10^{18}\ el/cm^3}$
($\approx {1.8\ 10^{-4}\ el/\Omega}$ where $\Omega$ is the unit cell).
As the laser transferred momentum is negligible, on the scale of 
the solid unit cell size, all pumped carriers are excited vertically from the valence
to the conduction bands along the the $\Gamma - L$ line.

In this configuration the e-e scattering channel is negligible.
Indeed the equilibrium e--e lifetimes are non zero only~\cite{Marini1}
for $\epsilon_{n\kk}-\epsilon_{CBM}>E_g$, with $E_g$ the energy gap, $\gee_{n\kk}$ the electronic energy and
$\epsilon_{CBM}$ the conduction band minimum\,(CBM).
In the case of the $L_1$ state  $\epsilon_{L_1}-\epsilon_{CBM}<E_g$ and, consequently, they are zero.
The additional NEQ contribution is due to inter--band processes. Its strength is linked to the
carrier density that, in the present case, is low enough to make the it
negligible~\cite{Bernardi1,Iveland1}.

{\em The two--photon photo--emission experiment, a gedanken experiment and the theoretical interpretation}.
In Fig.\ref{fig:1} the experimental occupation of the $L_1$ state (dots) is compared with the solution of Eq.\ref{eq:1} (continuous line).
The agreement between theory and experiment is excellent. Both the gradual filling  and emptying of the $L_1$
state follows quite nicely the experimental curve.
The theoretical results correctly describe the ultra--fast decay time--scale\,($\sim 180$\,fs) and
the $40\ fs$ shift of the population peak from the maximum of the pump pulse. 

The $40\ fs$ delay reflects the delicate balance
between the photo--excitation and the e--p scattering and can only be described by treating both
processes on the same footing.

Experimentally~\cite{Ichibayashi1,Ichibayashi2} the ultra--fast decay of the $L_1$ state is interpreted as due to
$L_1  \rightarrow X_1$ transitions. However a more deep 
analysis of the theoretical result reveals a different scenario. 

In Fig.\ref{fig:1}.$(a)$ the population of the levels at $t=0$ is shown.
Blue lines represent charges added and red lines charges removed. The band structure is
computed along the $L \Gamma L'$  high symmetry path in the Brillouin Zone\,(BZ).
In bulk $Si$ $L$ and $L'$ are equivalent points but  Fig.\ref{fig:1}.$(a)$ shows that
the level $L'_1$ is not populated and most of the carriers are injected in the $L_1$ level.

This symmetry breaking mechanism is made possible by the external field ($\UU$ operator in Eq.\ref{eq:1})
which, in the 2PPE experiment, is polarized along the crystallographic 
$\[111\]$ direction. This 
breaks the $L \leftrightarrow L'$ symmetry as the symmetry operation that moves $L$ in $L'$, although
being a symmetry of the unperturbed system, does not leave the $\[111\]$ direction unchanged.
In practice this means that Eq.\ref{eq:1} does not respect this symmetry anymore and
$\kk$--points connected by a rotation that does not leave the pumping field unchanged are populated
in a different way~\cite{Sachs1}.
Electrons are injected 
in the conduction band along the $\Gamma-L$ line but not, for symmetry reasons, along the $\Gamma-L'$
line.
This is clearly shown in Fig.~\ref{fig:1} where the population of the 
$L'_1$ state is represented with a dashed line.
The $L'_1$ state is gradually filled while $L_1$ is depleted revealing that the real source of the ultra--fast
decay observed experimentally is the $L_1\rightarrow L_1'$ scattering.

This scattering is faster than any other scattering as it involves states with the same energy.
Indeed Fig.\ref{fig:1} shows that initially the dynamics equilibrates the populations of the $L_1$ and $L'_1$ states which
reach the same value at $t\sim 220$\,fs. After this point both states decay simultaneously by using the slower
$L \rightarrow X$ channel towards the CBM~\cite{note_LX_dec}. 
After $t\sim 500$\,fs the electrons\,(holes) can be already be described by two Fermi distributions
around the CBM\,(VBM) with very high temperatures
($T_e\approx 9000\ K$ for electrons and $T_h\approx 2350\ K$ for hole).
Once the Fermi distributions are created, the relaxation process is mostly
dissipative and phonons are emitted in order to cool the 
carriers temperatures (after $t=1\ ps$, for example we obtain, $T_h\approx 550\ K$ and $T_e\approx 1900\ K$).

To better disentangle the $L_1\rightarrow L_1'$ process from the slower $L \rightarrow X$ channel
we consider a shorter laser pulse with $\sigma=50$\,fs. We also consider a {\em gedanken experiment}
where electrons are manually excited.
One of the approximations most widely used in the literature is to mimic the effect of the laser pulse
with some, {\em ad--hoc}, initial population of carriers in the valence bands. This approximation corresponds to
put $\UU={\bf 0}$ in Eq.\ref{eq:2} defining some initial arbitrary population $f_{n\kk}\(t=t_0\)$.
Here we have chosen 
an initial population around the $\Gamma_{15}$ state, with a carriers density
equal to the one measured experimentally. 

We then show (Fig.\ref{fig:1}.$(b)$) the population of the $L_1$ state in the {\em gedanken experiment} (dot--dashed line),
and when the photo--excitation is performed with the shorter pulse for both $L_1$ (continuous line) and of $L'_1$ (dashed line).
By comparing the results we notice that the decay of the $L_1$ state, when the carriers are manually excited is much
slower compared to the case when the carriers are photo--excited.
This is because the dynamics following the {\em ad--hoc} population is
symmetric and the $L_1$ and $L'_1$ are equally populated. The ultra--fast $L_1\rightarrow L_1'$ channel is switched off.
Instead, when the shorter pump pulse is considered, the two decay time--scales (the ultra--fast $L_1\rightarrow L_1'$ and 
the slower $L \rightarrow X$) are clearly visible. Immediately after the population peak we notice an almost vertical drop
of the $L_1$ population. The characteristic time scale is even faster ($90\ fs$) than the one measured experimentally
($180\ fs$). 
After the vertical drop we notice a more smooth
decay of both the $L_1$ and the $L'_1$ occupations, induced by the much slower $L \rightarrow X$ channel.

{\em Carrier lifetimes: an out--of--equilibrium concept}.  The most time--consuming part of
solving Eq.\ref{eq:1} is the update of the $\gc^{\(e,h\)}\(t\)$ functions whose dependence on the
occupations must be re--calculated at each time step.
A very tempting possibility would be to keep $\gc_{n\kk}^{\(e,h\)}\(t\)$ constant. This is indeed the main ingredient of the 
relaxation time approximation\,({\em RTA}) that is based on the assumption that  $\gc_{n\kk}^{\(e,h\)}\(t\)\sim\gc_{n\kk,eq}^{\(e,h\)}$, with $\gc_{n\kk,eq}^{\(e,h\)}$
the equilibrium lifetimes, calculated without the presence of any external field~\cite{EqNote}.
This approach has been recently used in 
Ref.~[\onlinecite{Bernardi1}] to describe the carrier relaxation in $Si$ excited by the weak sunlight. 

In order to investigate further the meaning of the $\gc_{n\kk}^{\(e,h\)}\(t\)$ time--dependence and their
crucial role in describing the  experimental results, let's introduce the totally--relaxed occupations, $f^{\infty}_{n\kk}$, 
defined as the occupations at the time $t_{rel}$ such that ${\partial_t f_{n\kk}\(t\)|_{t=t_{rel}}=0}$. 
From Eq.\ref{eq:1} and Eq.\ref{eq:3} it follows that
\begin{align} 
f^{\infty}_{n\mathbf{k}}\equiv f_{n\mathbf{k}}\(t_{rel}\)=\frac{\gamma_{n\mathbf{k}}^{(h)}\(t_{rel}\)}
     {\gamma_{n\mathbf{k}}^{(h)}\(t_{rel}\)+\gamma_{n\mathbf{k}}^{(e)}\(t_{rel}\)},
\label{eq:4}
\end{align}
where we have used the fact that at $t=t_{rel}$ the pump field is switched off and  $\left.\partial_t f_{n\kk}\(t_{rel}\)\right|_{pump}=0$.

In the equilibrium regime the dependence of the $\gc_{n\kk,eq}^{\(e/h\)}$ lifetimes on the electronic energies
is well--known~\cite{EqNote}.
Indeed for conduction bands $\gc_{n\kk,eq}^{\(e\)}=0$ while for valence states $\gc_{n\kk,eq}^{\(h\)}=0$. This means that,
from Eq.\ref{eq:4}, the totally--relaxed occupations would be zero for any conduction band. 

This is the reason why in the RTA~\cite{Bernardi1} $f^{\infty}_{n\mathbf{k}}$ is added as an adjustable parameter in the simulation.
It is commonly parametrized as a Fermi distribution with a given temperature and chemical potential. But our simulations reveal that, in general,
electrons and hole are distributed with 
two different Fermi distributions. This means two different chemical potentials and temperatures.
In our approach $f^{\infty}_{n\mathbf{k}}$ is a by--product of the simulation and it must not be provided at the
beginning. 

Moreover the present scheme allows to go well beyond the RTA by formally defining a {\em NEQ} carrier lifetime,
${\overline{\gamma}^{(e/h)}_{n\kk}\(t\)}$,
such that occupation functions satisfy the simple equation: ${\partial_t f^{(i)}=-\overline{\gamma}^{(i)} f^{(i)}}$ for ${i=e/h}$:
\begin{align} 
\overline{\gamma}^{\(e/h\)}_{n\kk}\(t\)=\frac{ \gamma_{n\kk}^{\(e\)}\(t\) f^{\(e\)}_{n\kk}\(t\)
- \gamma_{n\kk}^{\(h\)}\(t\) f^{\(h\)}_{n\kk}\(t\)  }{f^{\(e/h\)}_{n\kk}\(t\)} \text{.}
\label{eq:5}
\end{align}
Eq.\ref{eq:5} demonstrates that a true instantaneous NEQ carrier lifetime includes contributions from both the
electron ($\gamma_{n\kk}^{\(e\)}$) and the hole ($\gamma_{n\kk}^{\(h\)}$) lifetimes.

\begin{figure}[t]
\centering
\epsfig{figure=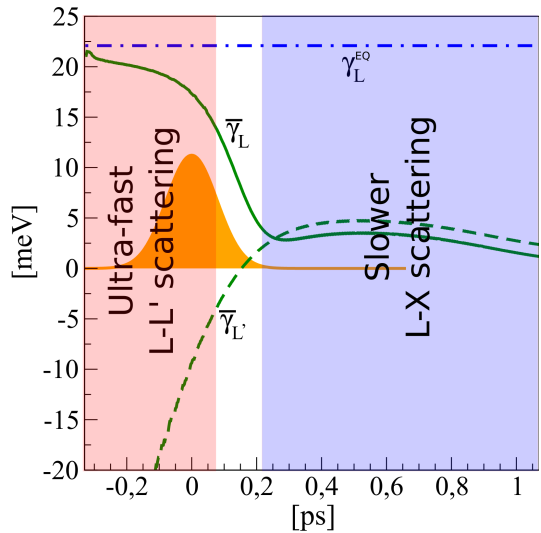,clip=,bbllx=0,bblly=0,bburx=400,bbury=400,width=0.4\textwidth}
\caption{\footnotesize{(color online) 
The equilibrium lifetime (blue dot-dashed line) is compared with the time dependent out--of--equilibrium lifetimes
defined in Eq.(\ref{eq:5}) for the $L_1$ and $L'_1$ states.
Due to the symmetry breaking induced by the laser pulse (the orange shadow represents its Gaussian envelope) we have two
in--equivalent lifetimes at $L_1$ (green line) and $L_1'$ (green dashed line). Their relative intensity define the ultra--fast
($\gamma_L \gg \gamma_{L'}$) and the slow ($\gamma_L \approx \gamma_{L'}$) time regimes.}}
\label{fig:3}
\end{figure}

The deviation of $\overline{\gamma}^{\(e/h\)}_{n\kk}\(t\)$ from $\gamma^{\(e/h\)}_{n\kk,eq}$ is, indeed, strictly connected
with the symmetry breaking mechanism that explains the experimental result. In the $L_1$/$L_1'$ case, indeed,
$\gamma_{L_1,eq}^{(e)}=\gamma_{L'_1,eq}^{(e)}\approx 20\ meV$. From Fig.\ref{fig:3} we see that, instead, both
$\overline{\gamma}^{\(e\)}_{L_1}\(t\)$ and $\overline{\gamma}^{\(e\)}_{L'_1}\(t\)$ changes, during the simulation, by an order
of magnitude. The reason is that, in Eq.\ref{eq:5}, the $\gamma^{(e)}_{n\kk}f^{(e)}_{n\kk}$ and 
$\gamma^{(h)}_{n\kk}f^{(h)}_{n\kk}$ factors are of the same order and their balance measures
the difference of population between the $L_1$ and $L'_1$ states. We can, therefore, 
easily recognize in Fig.\ref{fig:3} two well--defined regimes:
when $\overline{\gamma}^{(e)}_L \gg \overline{\gamma}^{(e)}_{L'}$ the ultra--fast $L_1 \rightarrow L'_1$ scattering channel is active. When
$\overline{\gamma}^{(e)}_L \approx \overline{\gamma}^{(e)}_{L'}$, instead, the relative $L_1$ and $L'_1$ populations are balanced and the dynamics
is dictated by the slower  $L\rightarrow X$ channel.

In conclusion we have presented a fully ab--initio simulation of the carrier dynamics in $Si$. 
The present scheme, based on  the merging of DFT with NEGF theory, successfully describes the ultra--fast
decay of the $L_1$ carrier population measured in a
recent 2PPE experiment. We have also highlighted that the microscopic mechanism that drives this ultra--fast decay 
is not a standard inter--valley scattering but it is due to an ultra--fast (as fast as $90$\,fs) 
$L_1\rightarrow L'_1$ scattering channel activated by the specific polarization the pump laser.
This physical interpretation is, also, supported by introducing a novel definition of the
non--equilibrium carrier lifetime that provides an intuitive picture of the 
physical processes activated by the initial photo--excitation. 

Financial support was provided by the Futuro in Ricerca grant No. RBFR12SW0J of the Italian
Ministry of Education, University and Research. D.~Sangalli would like to acknowledge G.~Onida
for the access granted to the etsfmi cluster in Milano and C.~Attaccalite for useful suggestions.

\bibliographystyle{apsrev4-1}
\bibliography{paper}

\begin{thebibliography}{33}%
\makeatletter
\providecommand \@ifxundefined [1]{%
 \@ifx{#1\undefined}
}%
\providecommand \@ifnum [1]{%
 \ifnum #1\expandafter \@firstoftwo
 \else \expandafter \@secondoftwo
 \fi
}%
\providecommand \@ifx [1]{%
 \ifx #1\expandafter \@firstoftwo
 \else \expandafter \@secondoftwo
 \fi
}%
\providecommand \natexlab [1]{#1}%
\providecommand \enquote  [1]{``#1''}%
\providecommand \bibnamefont  [1]{#1}%
\providecommand \bibfnamefont [1]{#1}%
\providecommand \citenamefont [1]{#1}%
\providecommand \href@noop [0]{\@secondoftwo}%
\providecommand \href [0]{\begingroup \@sanitize@url \@href}%
\providecommand \@href[1]{\@@startlink{#1}\@@href}%
\providecommand \@@href[1]{\endgroup#1\@@endlink}%
\providecommand \@sanitize@url [0]{\catcode `\\12\catcode `\$12\catcode
  `\&12\catcode `\#12\catcode `\^12\catcode `\_12\catcode `\%12\relax}%
\providecommand \@@startlink[1]{}%
\providecommand \@@endlink[0]{}%
\providecommand \url  [0]{\begingroup\@sanitize@url \@url }%
\providecommand \@url [1]{\endgroup\@href {#1}{\urlprefix }}%
\providecommand \urlprefix  [0]{URL }%
\providecommand \Eprint [0]{\href }%
\providecommand \doibase [0]{http://dx.doi.org/}%
\providecommand \selectlanguage [0]{\@gobble}%
\providecommand \bibinfo  [0]{\@secondoftwo}%
\providecommand \bibfield  [0]{\@secondoftwo}%
\providecommand \translation [1]{[#1]}%
\providecommand \BibitemOpen [0]{}%
\providecommand \bibitemStop [0]{}%
\providecommand \bibitemNoStop [0]{.\EOS\space}%
\providecommand \EOS [0]{\spacefactor3000\relax}%
\providecommand \BibitemShut  [1]{\csname bibitem#1\endcsname}%
\let\auto@bib@innerbib\@empty
\bibitem [{\citenamefont {O'Mara}\ \emph {et~al.}(1990)\citenamefont {O'Mara},
  \citenamefont {Herring},\ and\ \citenamefont {Hunt}}]{Omara1}%
  \BibitemOpen
  \bibfield  {author} {\bibinfo {author} {\bibfnamefont {W.}~\bibnamefont
  {O'Mara}}, \bibinfo {author} {\bibfnamefont {R.}~\bibnamefont {Herring}}, \
  and\ \bibinfo {author} {\bibfnamefont {L.}~\bibnamefont {Hunt}},\ }\href
  {http://books.google.it/books?id=COcVgAtqeKkC} {\emph {\bibinfo {title}
  {Handbook of Semiconductor Silicon Technology}}},\ Materials science and
  process technology series\ (\bibinfo  {publisher} {Noyes Publications},\
  \bibinfo {year} {1990})\BibitemShut {NoStop}%
\bibitem [{\citenamefont {Haug}\ and\ \citenamefont {Jauho}(2008)}]{Haug1}%
  \BibitemOpen
  \bibfield  {author} {\bibinfo {author} {\bibfnamefont {H.}~\bibnamefont
  {Haug}}\ and\ \bibinfo {author} {\bibfnamefont {A.-P.}\ \bibnamefont
  {Jauho}},\ }\href@noop {} {\emph {\bibinfo {title} {Quantum Kinetics in
  Transport and Optics of Semiconductors}}},\ edited by\ \bibinfo {editor}
  {\bibnamefont {Springer}}\ (\bibinfo {year} {2008})\BibitemShut {NoStop}%
\bibitem [{\citenamefont {Sundaram}\ and\ \citenamefont
  {Mazur}(2002)}]{Sundaram1}%
  \BibitemOpen
  \bibfield  {author} {\bibinfo {author} {\bibfnamefont {S.~K.}\ \bibnamefont
  {Sundaram}}\ and\ \bibinfo {author} {\bibfnamefont {E.}~\bibnamefont
  {Mazur}},\ }\href {\doibase 10.1038/nmat767} {\bibfield  {journal} {\bibinfo
  {journal} {Nat Mater}\ }\textbf {\bibinfo {volume} {1}},\ \bibinfo {pages}
  {217} (\bibinfo {year} {2002})}\BibitemShut {NoStop}%
\bibitem [{\citenamefont {Cerullo}\ and\ \citenamefont
  {De~Silvestri}(2003)}]{Cerullo1}%
  \BibitemOpen
  \bibfield  {author} {\bibinfo {author} {\bibfnamefont {G.}~\bibnamefont
  {Cerullo}}\ and\ \bibinfo {author} {\bibfnamefont {S.}~\bibnamefont
  {De~Silvestri}},\ }\href@noop {} {\bibfield  {journal} {\bibinfo  {journal}
  {Review of Scientific Instruments}\ }\textbf {\bibinfo {volume} {74}}
  (\bibinfo {year} {2003})}\BibitemShut {NoStop}%
\bibitem [{\citenamefont {Brida}\ \emph {et~al.}(2012)\citenamefont {Brida},
  \citenamefont {Manzoni}, \citenamefont {Cirmi}, \citenamefont {Polli},\ and\
  \citenamefont {Cerullo}}]{Brida1}%
  \BibitemOpen
  \bibfield  {author} {\bibinfo {author} {\bibfnamefont {D.}~\bibnamefont
  {Brida}}, \bibinfo {author} {\bibfnamefont {C.}~\bibnamefont {Manzoni}},
  \bibinfo {author} {\bibfnamefont {G.}~\bibnamefont {Cirmi}}, \bibinfo
  {author} {\bibfnamefont {D.}~\bibnamefont {Polli}}, \ and\ \bibinfo {author}
  {\bibfnamefont {G.}~\bibnamefont {Cerullo}},\ }\href {\doibase
  10.1109/JSTQE.2011.2125782} {\bibfield  {journal} {\bibinfo  {journal}
  {Selected Topics in Quantum Electronics, IEEE Journal of}\ }\textbf {\bibinfo
  {volume} {18}},\ \bibinfo {pages} {329} (\bibinfo {year} {2012})}\BibitemShut
  {NoStop}%
\bibitem [{\citenamefont {Shah}(1999)}]{Shah1}%
  \BibitemOpen
  \bibfield  {author} {\bibinfo {author} {\bibfnamefont {J.}~\bibnamefont
  {Shah}},\ }\href@noop {} {\emph {\bibinfo {title} {Ultrafast Spectroscopy of
  Semiconductors and Semiconductor Nanostructures}}},\ edited by\ \bibinfo
  {editor} {\bibnamefont {Springer}}\ (\bibinfo {year} {1999})\BibitemShut
  {NoStop}%
\bibitem [{\citenamefont {Hase}\ \emph {et~al.}(2003)\citenamefont {Hase},
  \citenamefont {Kitajima}, \citenamefont {Constantinescu},\ and\ \citenamefont
  {Petek}}]{Hase1}%
  \BibitemOpen
  \bibfield  {author} {\bibinfo {author} {\bibfnamefont {M.}~\bibnamefont
  {Hase}}, \bibinfo {author} {\bibfnamefont {M.}~\bibnamefont {Kitajima}},
  \bibinfo {author} {\bibfnamefont {A.~M.}\ \bibnamefont {Constantinescu}}, \
  and\ \bibinfo {author} {\bibfnamefont {H.}~\bibnamefont {Petek}},\ }\href
  {http://dx.doi.org/10.1038/nature02044} {\bibfield  {journal} {\bibinfo
  {journal} {Nature}\ }\textbf {\bibinfo {volume} {426}},\ \bibinfo {pages}
  {51} (\bibinfo {year} {2003})}\BibitemShut {NoStop}%
\bibitem [{\citenamefont {Hase}\ \emph {et~al.}(2012)\citenamefont {Hase},
  \citenamefont {Katsuragawa}, \citenamefont {Constantinescu},\ and\
  \citenamefont {Petek}}]{Hase2}%
  \BibitemOpen
  \bibfield  {author} {\bibinfo {author} {\bibfnamefont {M.}~\bibnamefont
  {Hase}}, \bibinfo {author} {\bibfnamefont {M.}~\bibnamefont {Katsuragawa}},
  \bibinfo {author} {\bibfnamefont {A.~M.}\ \bibnamefont {Constantinescu}}, \
  and\ \bibinfo {author} {\bibfnamefont {H.}~\bibnamefont {Petek}},\ }\href
  {http://dx.doi.org/10.1038/nphoton.2012.35} {\bibfield  {journal} {\bibinfo
  {journal} {Nat Photon}\ }\textbf {\bibinfo {volume} {6}},\ \bibinfo {pages}
  {243} (\bibinfo {year} {2012})}\BibitemShut {NoStop}%
\bibitem [{\citenamefont {Ichibayashi}\ \emph {et~al.}(2011)\citenamefont
  {Ichibayashi}, \citenamefont {Tanaka}, \citenamefont {Kanasaki},
  \citenamefont {Tanimura},\ and\ \citenamefont {Fauster}}]{Ichibayashi1}%
  \BibitemOpen
  \bibfield  {author} {\bibinfo {author} {\bibfnamefont {T.}~\bibnamefont
  {Ichibayashi}}, \bibinfo {author} {\bibfnamefont {S.}~\bibnamefont {Tanaka}},
  \bibinfo {author} {\bibfnamefont {J.}~\bibnamefont {Kanasaki}}, \bibinfo
  {author} {\bibfnamefont {K.}~\bibnamefont {Tanimura}}, \ and\ \bibinfo
  {author} {\bibfnamefont {T.}~\bibnamefont {Fauster}},\ }\href {\doibase
  10.1103/PhysRevB.84.235210} {\bibfield  {journal} {\bibinfo  {journal} {Phys.
  Rev. B}\ }\textbf {\bibinfo {volume} {84}},\ \bibinfo {pages} {235210}
  (\bibinfo {year} {2011})}\BibitemShut {NoStop}%
\bibitem [{\citenamefont {Ichibayashi}\ and\ \citenamefont
  {Tanimura}(2009)}]{Ichibayashi2}%
  \BibitemOpen
  \bibfield  {author} {\bibinfo {author} {\bibfnamefont {T.}~\bibnamefont
  {Ichibayashi}}\ and\ \bibinfo {author} {\bibfnamefont {K.}~\bibnamefont
  {Tanimura}},\ }\href {\doibase 10.1103/PhysRevLett.102.087403} {\bibfield
  {journal} {\bibinfo  {journal} {Phys. Rev. Lett.}\ }\textbf {\bibinfo
  {volume} {102}},\ \bibinfo {pages} {087403} (\bibinfo {year}
  {2009})}\BibitemShut {NoStop}%
\bibitem [{\citenamefont {Kwong}\ and\ \citenamefont {Bonitz}(2000)}]{Kwong1}%
  \BibitemOpen
  \bibfield  {author} {\bibinfo {author} {\bibfnamefont {N.-H.}\ \bibnamefont
  {Kwong}}\ and\ \bibinfo {author} {\bibfnamefont {M.}~\bibnamefont {Bonitz}},\
  }\href {\doibase 10.1103/PhysRevLett.84.1768} {\bibfield  {journal} {\bibinfo
   {journal} {Phys. Rev. Lett.}\ }\textbf {\bibinfo {volume} {84}},\ \bibinfo
  {pages} {1768} (\bibinfo {year} {2000})}\BibitemShut {NoStop}%
\bibitem [{\citenamefont {Jacoboni}\ and\ \citenamefont
  {Reggiani}(1983)}]{Jacoboni1}%
  \BibitemOpen
  \bibfield  {author} {\bibinfo {author} {\bibfnamefont {C.}~\bibnamefont
  {Jacoboni}}\ and\ \bibinfo {author} {\bibfnamefont {L.}~\bibnamefont
  {Reggiani}},\ }\href {\doibase 10.1103/RevModPhys.55.645} {\bibfield
  {journal} {\bibinfo  {journal} {Rev. Mod. Phys.}\ }\textbf {\bibinfo {volume}
  {55}},\ \bibinfo {pages} {645} (\bibinfo {year} {1983})}\BibitemShut
  {NoStop}%
\bibitem [{\citenamefont {Yabana}\ and\ \citenamefont
  {Bertsch}(1996)}]{K.1996}%
  \BibitemOpen
  \bibfield  {author} {\bibinfo {author} {\bibfnamefont {K.}~\bibnamefont
  {Yabana}}\ and\ \bibinfo {author} {\bibfnamefont {G.~F.}\ \bibnamefont
  {Bertsch}},\ }\href@noop {} {\bibfield  {journal} {\bibinfo  {journal} {Phys.
  Rev. B}\ }\textbf {\bibinfo {volume} {54}},\ \bibinfo {pages} {4484}
  (\bibinfo {year} {1996})}\BibitemShut {NoStop}%
\bibitem [{\citenamefont {Onida}\ \emph {et~al.}(2002)\citenamefont {Onida},
  \citenamefont {Reining},\ and\ \citenamefont {Rubio}}]{Onida1}%
  \BibitemOpen
  \bibfield  {author} {\bibinfo {author} {\bibfnamefont {G.}~\bibnamefont
  {Onida}}, \bibinfo {author} {\bibfnamefont {L.}~\bibnamefont {Reining}}, \
  and\ \bibinfo {author} {\bibfnamefont {A.}~\bibnamefont {Rubio}},\ }\href
  {\doibase 10.1103/RevModPhys.74.601} {\bibfield  {journal} {\bibinfo
  {journal} {Rev. Mod. Phys.}\ }\textbf {\bibinfo {volume} {74}},\ \bibinfo
  {pages} {601} (\bibinfo {year} {2002})}\BibitemShut {NoStop}%
\bibitem [{\citenamefont {Sato}\ \emph
  {et~al.}(2014{\natexlab{a}})\citenamefont {Sato}, \citenamefont {Yabana},
  \citenamefont {Shinohara}, \citenamefont {Otobe},\ and\ \citenamefont
  {Bertsch}}]{Sato1}%
  \BibitemOpen
  \bibfield  {author} {\bibinfo {author} {\bibfnamefont {S.~A.}\ \bibnamefont
  {Sato}}, \bibinfo {author} {\bibfnamefont {K.}~\bibnamefont {Yabana}},
  \bibinfo {author} {\bibfnamefont {Y.}~\bibnamefont {Shinohara}}, \bibinfo
  {author} {\bibfnamefont {T.}~\bibnamefont {Otobe}}, \ and\ \bibinfo {author}
  {\bibfnamefont {G.~F.}\ \bibnamefont {Bertsch}},\ }\href {\doibase
  10.1103/PhysRevB.89.064304} {\bibfield  {journal} {\bibinfo  {journal} {Phys.
  Rev. B}\ }\textbf {\bibinfo {volume} {89}},\ \bibinfo {pages} {064304}
  (\bibinfo {year} {2014}{\natexlab{a}})}\BibitemShut {NoStop}%
\bibitem [{\citenamefont {Sato}\ \emph
  {et~al.}(2014{\natexlab{b}})\citenamefont {Sato}, \citenamefont {Yabana},
  \citenamefont {Shinohara}, \citenamefont {Otobe},\ and\ \citenamefont
  {Bertsch}}]{Sato2}%
  \BibitemOpen
  \bibfield  {author} {\bibinfo {author} {\bibfnamefont {S.~A.}\ \bibnamefont
  {Sato}}, \bibinfo {author} {\bibfnamefont {K.}~\bibnamefont {Yabana}},
  \bibinfo {author} {\bibfnamefont {Y.}~\bibnamefont {Shinohara}}, \bibinfo
  {author} {\bibfnamefont {T.}~\bibnamefont {Otobe}}, \ and\ \bibinfo {author}
  {\bibfnamefont {G.~F.}\ \bibnamefont {Bertsch}},\ }\href {\doibase
  10.1103/PhysRevB.89.064304} {\bibfield  {journal} {\bibinfo  {journal} {Phys.
  Rev. B}\ }\textbf {\bibinfo {volume} {89}},\ \bibinfo {pages} {064304}
  (\bibinfo {year} {2014}{\natexlab{b}})}\BibitemShut {NoStop}%
\bibitem [{\citenamefont {Husser}\ and\ \citenamefont
  {Pehlke}(2012)}]{Husser1}%
  \BibitemOpen
  \bibfield  {author} {\bibinfo {author} {\bibfnamefont {H.}~\bibnamefont
  {Husser}}\ and\ \bibinfo {author} {\bibfnamefont {E.}~\bibnamefont
  {Pehlke}},\ }\href {\doibase 10.1103/PhysRevB.86.235134} {\bibfield
  {journal} {\bibinfo  {journal} {Phys. Rev. B}\ }\textbf {\bibinfo {volume}
  {86}},\ \bibinfo {pages} {235134} (\bibinfo {year} {2012})}\BibitemShut
  {NoStop}%
\bibitem [{\citenamefont {Bernardi}\ \emph {et~al.}(2014)\citenamefont
  {Bernardi}, \citenamefont {Vigil-Fowler}, \citenamefont {Lischner},
  \citenamefont {Neaton},\ and\ \citenamefont {Louie}}]{Bernardi1}%
  \BibitemOpen
  \bibfield  {author} {\bibinfo {author} {\bibfnamefont {M.}~\bibnamefont
  {Bernardi}}, \bibinfo {author} {\bibfnamefont {D.}~\bibnamefont
  {Vigil-Fowler}}, \bibinfo {author} {\bibfnamefont {J.}~\bibnamefont
  {Lischner}}, \bibinfo {author} {\bibfnamefont {J.~B.}\ \bibnamefont
  {Neaton}}, \ and\ \bibinfo {author} {\bibfnamefont {S.~G.}\ \bibnamefont
  {Louie}},\ }\href {\doibase 10.1103/PhysRevLett.112.257402} {\bibfield
  {journal} {\bibinfo  {journal} {Phys. Rev. Lett.}\ }\textbf {\bibinfo
  {volume} {112}},\ \bibinfo {pages} {257402} (\bibinfo {year}
  {2014})}\BibitemShut {NoStop}%
\bibitem [{\citenamefont {Attaccalite}\ \emph {et~al.}(2011)\citenamefont
  {Attaccalite}, \citenamefont {Gr\"uning},\ and\ \citenamefont
  {Marini}}]{Attaccalite1}%
  \BibitemOpen
  \bibfield  {author} {\bibinfo {author} {\bibfnamefont {C.}~\bibnamefont
  {Attaccalite}}, \bibinfo {author} {\bibfnamefont {M.}~\bibnamefont
  {Gr\"uning}}, \ and\ \bibinfo {author} {\bibfnamefont {A.}~\bibnamefont
  {Marini}},\ }\href {\doibase 10.1103/PhysRevB.84.245110} {\bibfield
  {journal} {\bibinfo  {journal} {Phys. Rev. B}\ }\textbf {\bibinfo {volume}
  {84}},\ \bibinfo {pages} {245110} (\bibinfo {year} {2011})}\BibitemShut
  {NoStop}%
\bibitem [{\citenamefont {Marini}(2013)}]{Marini1}%
  \BibitemOpen
  \bibfield  {author} {\bibinfo {author} {\bibfnamefont {A.}~\bibnamefont
  {Marini}},\ }\href {http://stacks.iop.org/1742-6596/427/i=1/a=012003}
  {\bibfield  {journal} {\bibinfo  {journal} {Journal of Physics: Conference
  Series}\ }\textbf {\bibinfo {volume} {427}},\ \bibinfo {pages} {012003}
  (\bibinfo {year} {2013})}\BibitemShut {NoStop}%
\bibitem [{\citenamefont {Marini}\ \emph {et~al.}(2009)\citenamefont {Marini},
  \citenamefont {Hogan},\ and\ \citenamefont {Varsano}}]{yambo_code}%
  \BibitemOpen
  \bibfield  {author} {\bibinfo {author} {\bibfnamefont {A.}~\bibnamefont
  {Marini}}, \bibinfo {author} {\bibfnamefont {M.}~\bibnamefont {Hogan},
  \bibfnamefont {C.~amd~Gr\"uning}}, \ and\ \bibinfo {author} {\bibfnamefont
  {D.}~\bibnamefont {Varsano}},\ }\href@noop {} {\bibfield  {journal} {\bibinfo
   {journal} {Computer Physics Communications}\ }\textbf {\bibinfo {volume}
  {180}},\ \bibinfo {pages} {1392} (\bibinfo {year} {2009})}\BibitemShut
  {NoStop}%
\bibitem [{\citenamefont {Giannozzi}\ \emph {et~al.}(2009)\citenamefont
  {Giannozzi}, \citenamefont {Baroni}, \citenamefont {Bonini}, \citenamefont
  {Calandra}, \citenamefont {Car}, \citenamefont {Cavazzoni}, \citenamefont
  {Davide}, \citenamefont {Chiarotti}, \citenamefont {Cococcioni},
  \citenamefont {Dabo}, \citenamefont {Corso}, \citenamefont {de~Gironcoli},
  \citenamefont {Fabris}, \citenamefont {Fratesi}, \citenamefont {Gebauer},
  \citenamefont {Gerstmann}, \citenamefont {Gougoussis}, \citenamefont
  {Kokalj}, \citenamefont {Lazzeri}, \citenamefont {Martin-Samos},
  \citenamefont {Marzari}, \citenamefont {Mauri}, \citenamefont {Mazzarello},
  \citenamefont {Paolini}, \citenamefont {Pasquarello}, \citenamefont
  {Paulatto}, \citenamefont {Sbraccia}, \citenamefont {Scandolo}, \citenamefont
  {Sclauzero}, \citenamefont {Seitsonen}, \citenamefont {Smogunov},
  \citenamefont {Umari},\ and\ \citenamefont {Wentzcovitch}}]{Giannozzi2009}%
  \BibitemOpen
  \bibfield  {author} {\bibinfo {author} {\bibfnamefont {P.}~\bibnamefont
  {Giannozzi}}, \bibinfo {author} {\bibfnamefont {S.}~\bibnamefont {Baroni}},
  \bibinfo {author} {\bibfnamefont {N.}~\bibnamefont {Bonini}}, \bibinfo
  {author} {\bibfnamefont {M.}~\bibnamefont {Calandra}}, \bibinfo {author}
  {\bibfnamefont {R.}~\bibnamefont {Car}}, \bibinfo {author} {\bibfnamefont
  {C.}~\bibnamefont {Cavazzoni}}, \bibinfo {author} {\bibfnamefont
  {C.}~\bibnamefont {Davide}}, \bibinfo {author} {\bibfnamefont {G.~L.}\
  \bibnamefont {Chiarotti}}, \bibinfo {author} {\bibfnamefont {M.}~\bibnamefont
  {Cococcioni}}, \bibinfo {author} {\bibfnamefont {I.}~\bibnamefont {Dabo}},
  \bibinfo {author} {\bibfnamefont {A.~D.}\ \bibnamefont {Corso}}, \bibinfo
  {author} {\bibfnamefont {S.}~\bibnamefont {de~Gironcoli}}, \bibinfo {author}
  {\bibfnamefont {S.}~\bibnamefont {Fabris}}, \bibinfo {author} {\bibfnamefont
  {G.}~\bibnamefont {Fratesi}}, \bibinfo {author} {\bibfnamefont
  {R.}~\bibnamefont {Gebauer}}, \bibinfo {author} {\bibfnamefont
  {U.}~\bibnamefont {Gerstmann}}, \bibinfo {author} {\bibfnamefont
  {C.}~\bibnamefont {Gougoussis}}, \bibinfo {author} {\bibfnamefont
  {A.}~\bibnamefont {Kokalj}}, \bibinfo {author} {\bibfnamefont
  {M.}~\bibnamefont {Lazzeri}}, \bibinfo {author} {\bibfnamefont
  {L.}~\bibnamefont {Martin-Samos}}, \bibinfo {author} {\bibfnamefont
  {N.}~\bibnamefont {Marzari}}, \bibinfo {author} {\bibfnamefont
  {F.}~\bibnamefont {Mauri}}, \bibinfo {author} {\bibfnamefont
  {R.}~\bibnamefont {Mazzarello}}, \bibinfo {author} {\bibfnamefont
  {S.}~\bibnamefont {Paolini}}, \bibinfo {author} {\bibfnamefont
  {A.}~\bibnamefont {Pasquarello}}, \bibinfo {author} {\bibfnamefont
  {L.}~\bibnamefont {Paulatto}}, \bibinfo {author} {\bibfnamefont
  {C.}~\bibnamefont {Sbraccia}}, \bibinfo {author} {\bibfnamefont
  {S.}~\bibnamefont {Scandolo}}, \bibinfo {author} {\bibfnamefont
  {G.}~\bibnamefont {Sclauzero}}, \bibinfo {author} {\bibfnamefont {A.~P.}\
  \bibnamefont {Seitsonen}}, \bibinfo {author} {\bibfnamefont {A.}~\bibnamefont
  {Smogunov}}, \bibinfo {author} {\bibfnamefont {P.}~\bibnamefont {Umari}}, \
  and\ \bibinfo {author} {\bibfnamefont {R.~M.}\ \bibnamefont {Wentzcovitch}},\
  }\href {http://stacks.iop.org/0953-8984/21/i=39/a=395502} {\bibfield
  {journal} {\bibinfo  {journal} {Journal of Physics: Condensed Matter}\
  }\textbf {\bibinfo {volume} {21}},\ \bibinfo {pages} {395502} (\bibinfo
  {year} {2009})}\BibitemShut {NoStop}%
\bibitem [{\citenamefont {R.M.Dreizler}\ and\ \citenamefont
  {E.K.U.Gross}(1990)}]{R.M.Dreizler1990}%
  \BibitemOpen
  \bibfield  {author} {\bibinfo {author} {\bibnamefont {R.M.Dreizler}}\ and\
  \bibinfo {author} {\bibnamefont {E.K.U.Gross}},\ }\href@noop {} {\emph
  {\bibinfo {title} {Density Functional Theory}}}\ (\bibinfo  {publisher}
  {Springer-Verlag},\ \bibinfo {year} {1990})\BibitemShut {NoStop}%
\bibitem [{\citenamefont {Sangalli}\ and\ \citenamefont
  {Marini}()}]{Sangalli1}%
  \BibitemOpen
  \bibfield  {author} {\bibinfo {author} {\bibfnamefont {D.}~\bibnamefont
  {Sangalli}}\ and\ \bibinfo {author} {\bibfnamefont {A.}~\bibnamefont
  {Marini}},\ }\href@noop {} {\enquote {\bibinfo {title} {Ab--initio
  out--of--equilibrium carriers dynamics},}\ }\BibitemShut {NoStop}%
\bibitem [{\citenamefont {Das~Sarma}\ \emph {et~al.}(1990)\citenamefont
  {Das~Sarma}, \citenamefont {Jain},\ and\ \citenamefont
  {Jalabert}}]{DasSarma1}%
  \BibitemOpen
  \bibfield  {author} {\bibinfo {author} {\bibfnamefont {S.}~\bibnamefont
  {Das~Sarma}}, \bibinfo {author} {\bibfnamefont {J.~K.}\ \bibnamefont {Jain}},
  \ and\ \bibinfo {author} {\bibfnamefont {R.}~\bibnamefont {Jalabert}},\
  }\href {\doibase 10.1103/PhysRevB.41.3561} {\bibfield  {journal} {\bibinfo
  {journal} {Phys. Rev. B}\ }\textbf {\bibinfo {volume} {41}},\ \bibinfo
  {pages} {3561} (\bibinfo {year} {1990})}\BibitemShut {NoStop}%
\bibitem [{\citenamefont {Allen}(1987)}]{Allen1}%
  \BibitemOpen
  \bibfield  {author} {\bibinfo {author} {\bibfnamefont {P.~B.}\ \bibnamefont
  {Allen}},\ }\href {\doibase 10.1103/PhysRevLett.59.1460} {\bibfield
  {journal} {\bibinfo  {journal} {Phys. Rev. Lett.}\ }\textbf {\bibinfo
  {volume} {59}},\ \bibinfo {pages} {1460} (\bibinfo {year}
  {1987})}\BibitemShut {NoStop}%
\bibitem [{bul()}]{bulk_prop}%
  \BibitemOpen
  \href@noop {} {}\bibinfo {note} {The time dependent occupation of the $L$
  points is interpreted, in the experiment~\cite{Ichibayashi1,Ichibayashi2}, as
  a bulk property.}\BibitemShut {Stop}%
\bibitem [{\citenamefont {Iveland}\ \emph {et~al.}(2013)\citenamefont
  {Iveland}, \citenamefont {Martinelli}, \citenamefont {Peretti}, \citenamefont
  {Speck},\ and\ \citenamefont {Weisbuch}}]{Iveland1}%
  \BibitemOpen
  \bibfield  {author} {\bibinfo {author} {\bibfnamefont {J.}~\bibnamefont
  {Iveland}}, \bibinfo {author} {\bibfnamefont {L.}~\bibnamefont {Martinelli}},
  \bibinfo {author} {\bibfnamefont {J.}~\bibnamefont {Peretti}}, \bibinfo
  {author} {\bibfnamefont {J.~S.}\ \bibnamefont {Speck}}, \ and\ \bibinfo
  {author} {\bibfnamefont {C.}~\bibnamefont {Weisbuch}},\ }\href {\doibase
  10.1103/PhysRevLett.110.177406} {\bibfield  {journal} {\bibinfo  {journal}
  {Phys. Rev. Lett.}\ }\textbf {\bibinfo {volume} {110}},\ \bibinfo {pages}
  {177406} (\bibinfo {year} {2013})}\BibitemShut {NoStop}%
\bibitem [{\citenamefont {Sachs}(1957)}]{Sachs1}%
  \BibitemOpen
  \bibfield  {author} {\bibinfo {author} {\bibfnamefont {M.}~\bibnamefont
  {Sachs}},\ }\href {\doibase 10.1103/PhysRev.107.437} {\bibfield  {journal}
  {\bibinfo  {journal} {Phys. Rev.}\ }\textbf {\bibinfo {volume} {107}},\
  \bibinfo {pages} {437} (\bibinfo {year} {1957})}\BibitemShut {NoStop}%
\bibitem [{not()}]{note_LX_dec}%
  \BibitemOpen
  \href@noop {} {}\bibinfo {note} {The decay time of our simulated $L_1$
  population matches exactly the experiment up to ${\approx t=200\ fs}$. Only
  at later times we see a small deviation from the measured behavior which,
  however, corresponds to the activation of the slower $L_1\rightarrow X_1$
  processes. Experimentally the laser is shone on a surface which can penetrate
  the sample, at the experimental wave--length, for $\approx 10\ nm$. Thus the
  electrons have access to a large number of of empty states degenerate with
  $L_1$. Also surface state close in energy may exist. This explains the small
  deviation of our simulation from the experimental result.}\BibitemShut
  {Stop}%
\bibitem [{EqN()}]{EqNote}%
  \BibitemOpen
  \href@noop {} {}\bibinfo {note} {The equilibrium lifetimes are calculated
  within the $GW$ approximation~\cite{0034-4885-61-3-002} for the e--e channel
  and within the Fan approximation~\cite{Fan1950} for the e--p channel. The
  non--equilibrium lifetimes are obtained by extending to non--equilibrium
  regime the $GW$ and Fan approximations as described in
  Ref.\onlinecite{Marini1}.}\BibitemShut {Stop}%
\bibitem [{\citenamefont {Aryasetiawan}\ and\ \citenamefont
  {Gunnarsson}(1998)}]{0034-4885-61-3-002}%
  \BibitemOpen
  \bibfield  {author} {\bibinfo {author} {\bibfnamefont {F.}~\bibnamefont
  {Aryasetiawan}}\ and\ \bibinfo {author} {\bibfnamefont {O.}~\bibnamefont
  {Gunnarsson}},\ }\href {http://stacks.iop.org/0034-4885/61/i=3/a=002}
  {\bibfield  {journal} {\bibinfo  {journal} {Reports on Progress in Physics}\
  }\textbf {\bibinfo {volume} {61}},\ \bibinfo {pages} {237} (\bibinfo {year}
  {1998})}\BibitemShut {NoStop}%
\bibitem [{\citenamefont {Fan}(1950)}]{Fan1950}%
  \BibitemOpen
  \bibfield  {author} {\bibinfo {author} {\bibfnamefont {H.~Y.}\ \bibnamefont
  {Fan}},\ }\href@noop {} {\bibfield  {journal} {\bibinfo  {journal} {Phys.
  Rev.}\ }\textbf {\bibinfo {volume} {78}},\ \bibinfo {pages} {808} (\bibinfo
  {year} {1950})}\BibitemShut {NoStop}%
\end{thebibliography}%

\end{document}